# Vectorization of Gradient Boosting of Decision Trees Prediction in the CatBoost Library for RISC-V Processors

Evgeny Kozinov[1][0000-0001-6776-0096], Evgeny Vasiliev[1][0000-0002-7949-1919],
Andrey Gorshkov, Valentina Kustikova[1][0000-0002-6159-1145], Artem Maklaev,
Valentin Volokitin[1][0000-0003-1075-1329], Iosif Meyerov[1][0000-0001-6905-2050]

[1] Lobachevsky State University of Nizhni Novgorod, 603950 Nizhni Novgorod, Russia
`meerov@vmk.unn.ru`

**Abstract.** The emergence and rapid development of the open RISC-V instruction set architecture opens up new horizons on the way to efficient devices, ranging from existing low-power IoT boards to future high-performance servers. The effective use of RISC-V CPUs requires software optimization for the target platform. In this paper, we focus on the RISC-V-specific optimization of the CatBoost library, one of the widely used implementations of gradient boosting for decision trees. The CatBoost library is deeply optimized for commodity CPUs and GPUs. However, vectorization is required to effectively utilize the resources of RISC-V CPUs with the RVV 0.7.1 vector extension, which cannot be done automatically with a C++ compiler yet. The paper reports on our experience in benchmarking CatBoost on the Lichee Pi 4a, RISC-V-based board, and shows how manual vectorization of computationally intensive loops with intrinsics can speed up the use of decision trees several times, depending on the specific workload. The developed codes are publicly available on GitHub.

**Keywords:** RISC-V, Gradient Boosting Trees, Decision Trees, Machine Learning, Performance Analysis, Performance Optimization, Vectorization.

## 1 Introduction

An open RISC-V instruction set architecture (ISA) was originally developed at the University of California at Berkeley, and has rapidly spread throughout the world. The openness and extensibility of the architecture, the lack of patent royalties, the simplicity, brevity and consistency of the RISC-V ISA have received well-deserved recognition in the academic community and high-tech industry. Today, there is a rapid increase in the interest in RISC-V, expressed in the development of new ISA extensions, the release of RISC-V-based devices, planning of new projects to create RISC-V clusters, and investments in the development of relevant educational materials. Now we cannot confidently say that RISC-V will take a place among computing architectures on a par with x86 and ARM, but the current state looks promising.

Our HPC laboratory works in the field of researching and development of methods for improving the performance of numerical simulation software. Understanding the promise of RISC-V, we began working with early publicly available boards to estab-lish how to evaluate performance and what techniques are applicable to improve performance on RISC-V CPUs. We started with standard benchmarks [1] and gained intuition about how to optimize small programs to utilize resources of RISC-V CPUs. In this paper we take the next step in this direction. In particular, we look at CatBoost [2], one of the common libraries for solving machine learning (ML) problems using decision trees. We consider CatBoost algorithms as a "black box", without delving too deeply into the logic of the algorithms. We focus on the performance aspects, and look for ways to speed up the code by detecting and vectorizing computationally intensive loops along the critical paths of program execution on different workloads.

The choice of CatBoost as a testbed is due to the fact that it is a fairly popular, complex, and large framework. We demonstrate how to speed up a large framework for RISC-V CPUs using manual vectorization, without going into details of how the algorithms work or having a fully functional set of tools such as VTune [3], Advisor [4] or likwid [5]. The scientific contribution of the paper is as follows:

1. We describe a method for identifying performance bottlenecks in the absence of developed tools and with the complex architecture of the target application.
2. Using CatBoost as an example, we demonstrate how to accelerate main loops for RISC-V through manual vectorization where the compiler cannot handle it yet.
3. Overall, we show that even the current capabilities of RISC-V devices enable to solve quite complex problems and it makes sense to address performance issues now, without waiting for the release of more powerful devices.

## 2    Related Work

In recent years, the scientific community has been actively exploring the potential of the RISC-V architecture. Most papers focus on the development of the ISA, with less attention being paid to optimizing software performance on RISC-V CPUs. However, with the prospect of high-performance RISC-V CPUs entering the market, bridging the gap between HPC community and RISC-V has become an increasingly popular topic. This is evident in the announcement of RISC-V workshops at major HPC conferences such as ISC High Performance, HPC Asia, EuroPar, and PPAM.

At the moment, not many studies have been published regarding performance analysis and software optimization on RISC-V CPUs. However, an overview of RISC-V vector extensions and the corresponding infrastructure is given in [6]. Paper [7] contains a performance evaluation of HPC workloads using FPGA simulation. The paper [8] focuses on the important problem of insufficient development of compilers for RISC-V in terms of code vectorization. Indeed, current compilers only target the RVV 1.0 vector extensions, while most of the available devices support only RVV 0.7.1. The paper proposes an approach that involves automatic translation from the RVV 1.0 into RVV 0.7.1 instruction set. The paper [9] proposes enhancements to a RISC-V VPU to improve the efficiency of sparse algebra operations, which is a bottleneck in many scientific simulations. Another approach to improve the performance of sparse algebra algorithms, in particular, Sparse Matrix-Vector Multiplication (SpMV), is proposed in [10]. The authors conclusively demonstrate the possibil-ity of modeling the performance of SpMV on RISC-V CPUs using the Roofline Model and propose an approach to increasing the efficiency of hardware usage based on software-hardware co-design. Paper [11] explores another mathematical kernel, Fast Fourier Transform (FFT). It is demonstrated that the use of RISC-V-specific optimizations can significantly speed up calculations on RISC-V CPUs. In [12], the authors present comprehensive benchmarking results of OpenFOAM, one of the most widely used frameworks for scientific simulations. Their findings compare the performance and power consumption across devices of various architectures, including the Nezha D1 RISC-V board. In [13] we demonstrated how several image processing algorithms from OpenCV could be accelerated through more efficient vectorization for RISC-V. The results of various researchers indicate that RISC-V CPUs have promise for HPC and there is an urgent need for new ideas both in hardware development and in methods for optimizing software using specific applications in demand.

In this paper, we consider CatBoost [2], a high-performance library that is used by researchers to solve problems using ML methods. When working with large volumes of data, the time to train and apply trained models becomes critical in many areas of application. Therefore, the authors of CatBoost present an analysis of model training and application performance [14], as well as comparisons with existing decision tree ensembles implementations from XGBoost [15] and LightGBM [16] with similar parameters. It is shown [14] that CatBoost is ahead of its analogues on the same test infrastructure. The performance of CatBoost algorithms has also been studied in other projects. One of the most comprehensive performance studies [17] reports an in-depth analysis and comparison of the quality and training speed of various implementations of ML methods based on decision trees using a large number of publicly available datasets. Along with model training performance, reducing prediction time is also of interest. The paper [18] shows the advantages in terms of time and memory consumption of CatBoost compared to Random Forest and Support Vector Machine methods.

In general, the CatBoost framework has been carefully optimized for various architectures, including both data structures and algorithms as well as the low-level optimizations of the main computational kernels. However, there is still a lack of RISC-V-specific optimizations, and this research has been conducted in this direction.

## 3    CatBoost Overview

CatBoost is a machine learning algorithm which is developed by Yandex researchers and engineers. It is used in search and recommendation systems, personal assistants, self-driving cars software and in many other applications, in particular, at Yandex, CERN, Cloudflare, Careem taxi [2]. It is available as an open-source library distributed under the Apache 2.0 license [19]. This library supports ranking, classification, regression and other machine learning tasks [20]. There are Python, R, Java and C++ programming interfaces. CatBoost supports CPU and GPU computations. The library also includes tools for training and applying trained models, analyzing the quality, as well as visualization tools [2].

CatBoost [14, 21] is based on the gradient boosting of decision trees. The idea of boosting is to build a strong predictive model using an ensemble of weaker ones. This algorithm uses *oblivious decision trees* with a small depth as weak models. An oblivious tree is a simplified tree model in which every node at the same level checks the same branching condition. During training, trees are added to the ensemble sequentially and each tree attempts to correct the errors of the previous one.

Applying CatBoost starts from encoding features to floating point values. Further, the feature descriptor of each input sample is binarized. The computed binarized features are used to calculate predictions as follows. For each input feature vector, a tree traversal is performed from root to leaf, guided by checking conditions at each node of the tree. CatBoost implementation optimizes the traversal using bitwise operators to eliminate branching. The final decision is made based on combining the results obtained from each tree. Let's take a closer look at the procedure. An oblivious decision tree contains exactly $2^{d-1}$ leafs, where $d$ is the depth of the tree. In such models, the index of each leaf is encoded by a binary vector with a length equal to the depth of the tree. In this vector, zero corresponds to the left child (a condition at a tree node is False), and one – to the right child (a condition is True). The leaf number is determined by the set of binarized feature values that split the tree at the levels. The indexes of leaves in a binary representation can be calculated as $\sum_{i=0}^{d-1} 2^i B(x, f(t,i))$, where $i$ is the depth at which the current node is located, $t$ is the index of the tree in the ensemble, $f(t,i)$ is the index of the binarized feature in the descriptor by which split is performed in the tree with index $t$ at depth $i$, and $B(x, f(t,i))$ is the binary function representing the split result. This approach accelerates prediction for each tree in the ensemble and increases overall efficiency.

The CatBoost library provides better quality metrics than many state-of-the-art algorithms on a wide range of datasets [21]. The significant advantage of CatBoost is its ability to automatically process both numeric and categorical features. As previously mentioned, the library is highly optimized and demonstrates high performance in model training and testing on commodity CPUs and GPUs [14].

## 4 Performance Optimization for RISC-V Architecture

In this section we present a high-level description of the benchmarking and optimization methodology employed in the paper. We describe datasets, discuss RISC-V-specific optimization opportunities and profiling issues, present our benchmarking methodology, and show how main hotspots have been identified and vectorized.

### 4.1 Benchmarks

The *Covertype* dataset [24] contains 52 integer and binary features representing wilderness areas and soil types, needed to predict forest cover type (7 classes). The dataset was randomly split into a training set and a testing set in a 70:30 ratio. *The Santander customer transaction dataset* [26] contains 200 non-normalized features along with a binary target variable. The train and test parts have 200000 samples each. The

*YearPredictionMSD* dataset [25] contains 90 non-normalized features extracted from songs, needed to predict the year in which the song was released. The dataset is split into 463715 train samples and 51630 test samples according to the dataset developers idea. MQ2008 dataset [27] contains 46 features for solving a supervised ranking task. The dataset contains 9630 train and 2874 test samples. The *image-embeddings* dataset is a subset of the *PASCAL VOC 2007* dataset [28] that contains only images of one object class out of twenty possible. Train and test parts include 2808 and 2841 images respectively. Embeddings were generated for images using the pre-trained *resnet34* model from the TorchVision library. To obtain the embeddings, the last classification layer was removed from the model. The parameters of the Catboost models are shown in Table 1.

**Table 1.** Datasets for performance analysis and optimization and their parameters. The maximum number of training iterations is set to 10000. Other parameters are set to default values.

| Dataset | Rows x Cols | # of Classes | LossFunction | Learning rate | Tree depth |
|---|---|---|---|---|---|
| MQ2008 | 9630 x 46 | - | YetiRank | 0.02 | 6 |
| Santander customer transaction | 400k x 202 | 2 | LogLoss | 0.01 | 1 |
| Covertype | 464.8k x 54 | 7 | MultiClass | 0.50 | 8 |
| YearPredictionMSD | 515k x 90 | - | MAE | 0.30 | 6 |
| image-embeddings | 5649 x 512 | 20 | MultiClass | 0.05 | 4 |

### 4.2 Optimization Opportunities

Analyzing and optimizing the performance of a large open-source framework is a challenging task. Such frameworks are often created over many years by a large, distributed community of developers. Even a high-level study of such a project by third-party developers with immersion in the implementation details requires a lot of effort and does not always lead to success. Note also that on traditional architectures, we have powerful optimizing compilers and profilers. Regarding RISC-V, the work is much more complicated, as the capabilities of the software infrastructure are still quite limited. In this regard, we decided not to go into the algorithmic details but to limit ourselves to local optimization of the main *hotspots*. Thus, *vectorization* of computationally-intensive loops is one of the promising methods to speed up the calculations. Considering that current versions of compilers for RISC-V do not allow automatic vectorization of codes for currently available RISC-V CPUs with the RVV 0.7.1 vector instruction set, and sufficiently powerful devices that support RVV 1.0 are not publicly available yet, we have focused on manual vectorization using intrinsics. However, in order to implement this approach we need to identify the main hotspots within a fairly large codebase of the CatBoost library.

### 4.3 How to Find Hotspots?

The methodology for performance analysis and optimization has been known for many years and has been quite well developed. However, when analyzing the performance of CatBoost on RISC-V boards, we encountered a number of problems. As mentioned above, the community has not yet developed such powerful profilers like those for RISC-V CPUs as for x86 and ARM CPUs. We found that using the *perf* tool from the Linux OS to profile and analyze CatBoost is problematic, as it consists of a Python interface and algorithms implemented in a *dynamic libraries* in C++. Therefore, we had to develop custom profiling primitives to determine the main hotspots, build a call graph, and find critical paths in the call graphs that arise in prediction runs on specific datasets.

We use the following way to analyze the performance of the CatBoost prediction. The first stage in profiling is to search for the entry point in the ML prediction algorithms. In this regard we use perf tool. In particular, we found that calling the `ApplyModelMulti()` function takes up the majority of the time during prediction. This function is the entry point for solving a classification problem with several categories on a test dataset. At the next stage, we analyzed the individual function calls that occur during CatBoost prediction runs on selected datasets, and formulated assumptions regarding the main hotspots. To test the assumptions and accurately measure the running time of functions, we developed a relatively simple C++ class. This class allowed us to inject time measurements into the code and combine the results to account for multiple of function calls. For several modes, we were able to identify the main hotspots, build call graphs and determine the critical paths in the function calling scheme.

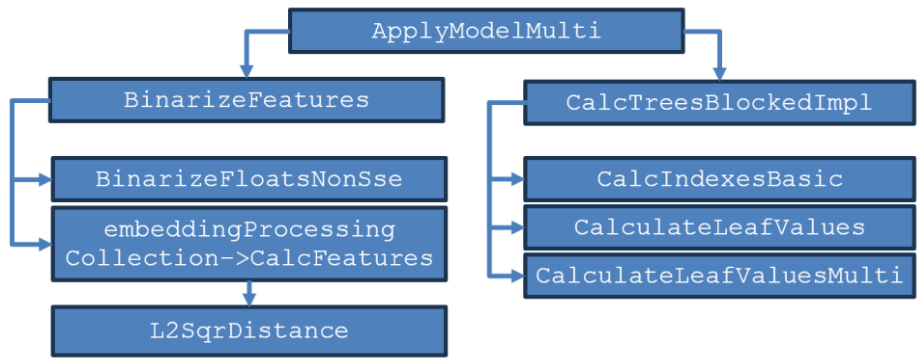

**Fig. 1.** Simplified call graph for main hotspots

The example of a simplified call graph for several datasets is shown in Fig. 1. According to our analysis, for all models corresponding datasets from Table 1, except for "image-embeddings", the processing of the decision tree stored in the model takes most of the time. In this case, the `CalcTreesBlockedImpl()` function is the main hotspot. This function calls the `CalcIndexesBasic()` and `CalculateLeafValues()` or `CalculateLeafValuesMulti()` functions depending on the number of predicted classes. The `CalcIndexesBasic()` function determines the leaf index which we obtain when traversing each tree during testing. `CalculateLeafValues()` и `CalculateLeafValuesMulti()` calculate the value at the corresponding leaf for binary or multi-class classification, respectively. The second most important candidate for optimization is the `BinarizeFeatures()` function, which calls the `BinarizeFloatsNonSse()` function. This function is responsible for the binarizing of float features for each sample in the test dataset. For the "image-embeddings" dataset, the main calculation time is spent on feature extraction using the KNN algorithm [22]. In this algorithm, most of the time is spent searching for neighbors, which in turn requires calculating distances using the L2 norm. The `L2SqrDistance()` function is an obvious candidate for optimization.

Note that the functions corresponding to the leaves of the call graph (Fig. 1) are called from hundreds (in the case of `BinarizeFloatsNonSse()`) to hundreds of thousands of times (in the case of `L2SqrDistance()`). In this case, the object-oriented approach previously discussed can be used to find hotspots, but it introduces a very large profiling overhead. To solve the problem, we measure the execution time of functions by accumulating it into simple C-style variables.

The results of profiling and optimizing the code have been extensively validated through several runs. Firstly, we compared the accuracy of the values achieved on x86 CPUs with those obtained on RISC-V CPUs. Secondly, we analyzed the overhead of measuring computation time during profiling by comparing it with solving a problem without profiling. For our initial assessment, we used reduced datasets of 1000 samples, but our final conclusions were based on the full datasets from Table 1. Overall, the information we obtained regarding the main hotspots and their contribution to the overall time has allowed us to optimize CatBoost for RISC-V processors.

### 4.4 Vectorization of Hotspots

The current implementation of CatBoost is not vectorized for RISC-V processors, which does not allow efficient utilization of RISC-V CPUs efficiently. Therefore, we decided to vectorize the main hotspots using RVV 0.7.1 intrinsics. We discuss optimizations for each of the main hotspots below.

**CalcIndexesBasic().** In this function it is necessary to vectorize a loop shown in Fig. 2 (left). The body of this loop includes the following operations. The value of the binarized feature of an input sample is compared to a threshold at each level of the tree. If the value is greater than or equal to the threshold, the corresponding bit in the resulting array is set to one.

The method of vectorization is shown in Fig. 2 (right). Firstly, we prepare a vector of ones using the vector integer move intrinsic `vmv_v_x_u32m4()` and shift each element by the number of positions corresponding to the level of the tree using the vector bit shift intrinsic `vsll_vx_u32m4()`. Then, in the loop, a binary mask is formed by comparing the features with a threshold value (see Fig. 2, where green and orange colors correspond to the True and False values, respectively). In this regard, the intrinsic `vmsgeu_vx_u8m1_b8()` is used. The resulting mask is then used to perform a bitwise `or` operation between the vector of units and the resulting vector (intrinsic `vor_vv_u32m4_m()`). Compared to the baseline, the shift operation is performed once before the main loop, which also reduces the number of operations.

**CalculateLeafValues() and CalculateLeafValuesMulti().** Both functions perform vector addition with indirect addressing. It is known that vectorization of such loops requires the use of time-consuming instructions like scatter/gather, which implement non-unit stride access to memory. The RVV 0.7.1 instruction set contains such operations, but they have a significant overhead, which in this case does not pay off due to the very small number of arithmetic operations with data. However, we hope that future CPUs of the RISC-V architecture will execute such vector codes more efficiently, as happened before with Intel x86 CPUs.

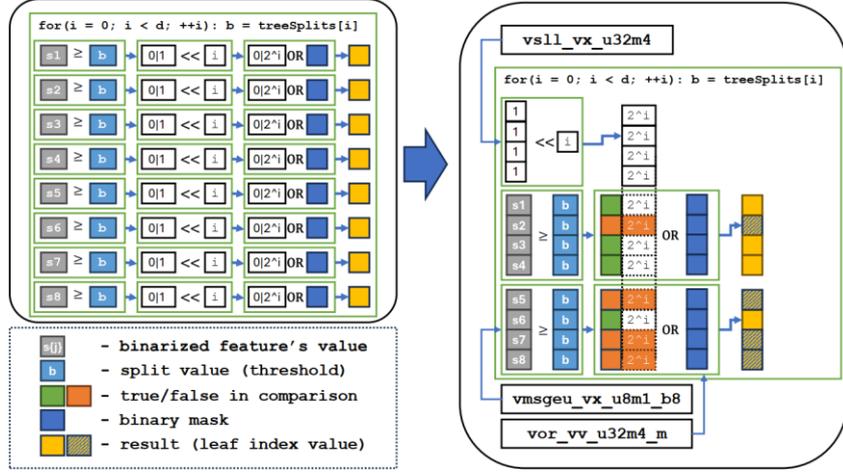

**Fig. 2.** Vectorization of the `CalcIndexesBasic()` function. Scheme of calculations in the original loop (left panel) and in the vectorized loop (right panel)

**BinarizeFloatsNonSse().** This function performs feature binarization. In this regard, floating point feature values are divided into bins, the boundaries of which are determined at the training stage. As a result, an integer bin index is assigned to each feature. In this algorithm (see Fig. 3, left) the inner loop is computationally intensive. We vectorize it as follows. Firstly, we unrolled the outer loop over feature values and then replaced the operations from the inner loop with their vectorized versions (see Fig. 3, right). At the beginning, a vector of ones is created (`vmv_v_x_u8m1()`). Next, in the loop, values are loaded into a vector register for comparison to the boundaries of bins. They are compared (`vmfgt_vf_f32m4_b8()`) and a mask is calculated. Then the unit vector is added to the resulting vector, taking into account the mask (`vadd_vv_u8m1_m()`). Finally, the accumulated bin indexes are stored in a resulting array.

**L2SqrDistance().** The function is used to determine the distance between vectors based on the L2 norm and is used often in CatBoost algorithms. Fig. 4 (left) shows the scalar implementation. Vectorization of this function is shown in Fig. 4 (right). Firstly, we set all the vector variables that correspond to the partial sums to zero. Then, in a loop, each part of the processed vector is loaded into vector variables. After that, the squares of the differences between elements of the vectors are calculated using (`vfsub_vv_f32m4()`), and the results are added to the vector variable (`vfmacc_vv_f32m4()`). At the end of the loop, the vector variable contains the partial sums for the L2 distance. Next, reduction is performed using the intrinsic function `vfredsum_vs_f32m4_f32m1()`.

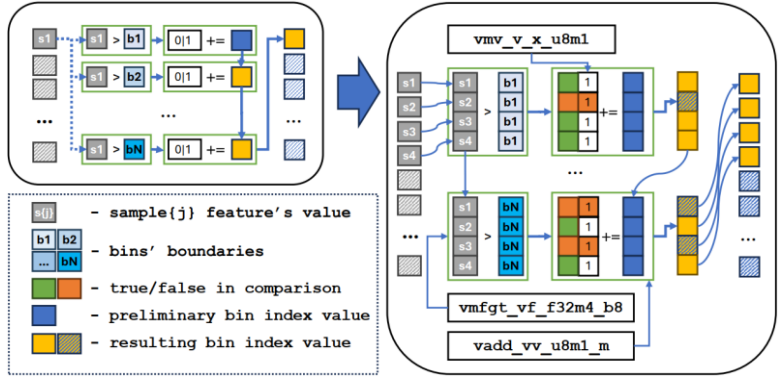

**Fig. 3.** Vectorization of the `BinarizeFloatsNonSse()` function. Schemes of calculations in the original loop (left panel) and in the vectorized loop (right panel)

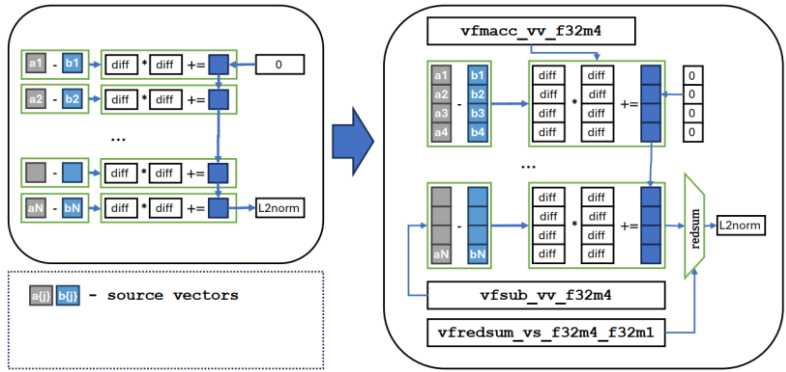

**Fig. 4.** Vectorization of the `L2SqrDistance()` function. Schemes of calculations in the original loop (left panel) and in the vectorized loop (right panel)

Note that the vector extension RVV 0.7.1 allows the adjustment of the number of vector registers used during operations. For example, we could use a block of four 128-bit registers along with the corresponding blockwise vector operations to process data as if the architecture contained 512-bit registers. This method corresponds to the `m4` suffix in the names of intrinsic functions. In practice, this method can significantly improve performance, but determining the best mode (`m1`, `m2`, `m4`, `m8`) requires experiments. Our implementation uses the option that yields maximum performance.

## 5 Numerical Results

### 5.1 Hardware

Computational experiments were performed on the Lobachevsky supercomputer with the following equipment:

1. Computational nodes based on Intel Xeon Silver 4310T (2 CPUs with 10 cores each, 20 cores in total), 64 GB RAM.
2. Mini-cluster Lichee Cluster 4A [23]. The mini-cluster includes 7 boards with RISC-V TH1520 CPUs based on C910 cores. Each processor contains 4 cores with support for RVV 0.7.1 vector extensions. Each board has 16GB of RAM available.

Intel CPUs were used to train models, prepare datasets, compute test accuracy, and compare performance. For x86 CPUs, CatBoost is built according to the guidelines without any changes of the source code and build scripts. To simplify the deployment of the library, a ready-made assembly from conda is used. However, for RISC-V CPUs, we build CatBoost from source codes in two stages. Due to a lack of native compiler support for vector extensions at the time of development, as the first step in resolving dependencies, we compiled the functions with RVV 0.7.1 intrinsics with the gcc-8.4.0 cross-compiler as a static library. At the second stage, the main package was built directly on a RISC-V board using the clang 14.0.6 compiler. Dependencies on vector functions were resolved from an already-built library of optimized functions.

### 5.2 Performance Analysis on x86 and RISC-V CPUs

Performance and correctness testing can be done by comparing the results of CatBoost running on x86 CPUs (code optimized by the CatBoost developers) and RISC-V CPUs (baseline scalar code from the CatBoost developers and our vectorized implementation). We split all experiments into two groups. *The first group of tests* was carried out on reduced datasets of 1000 samples and was used for initial correctness assessment and performance evaluation during development. All such runs were done in a *serial mode* to simplify accurate time measurements. Given the relatively small sample size, the code ran on RISC-V CPUs for a reasonable amount of time and allowed for profiling and experimentation with the code while continuously monitoring quality metrics. *The second group of experiments* involved large-scale runs in a *multi-threaded mode* on full sets of samples and was used for the final assessment.

First, consider the results for 1000 samples. The prediction was perfomed simultaneously for all the samples. Tables 2-4 contain the experimental results for three datasets, namely, *YearPredictionMSD*, *Covertype*, and *image-embedding*, respectively. The *YearPredictionMSD* dataset is used to solve a regression problem, while the *Covertype* and *image-embedding* datasets are used for multi-class classification. Unlike *YearPredictionMSD* and *Covertype*, the *image-embedding* dataset requires additional data preprocessing to extract features. Note that these datasets show significant differences in terms of CatBoost's performance profile. As can be seen in Tables 2 and 3, processing of a decision tree is the most time-consuming part of the prediction phase.

When predicting a single class, significant amount of time is taken by calling the `CalcIndexesBasic()` function. Calls of the `CalculateLeafValues()` and `BinarizeFloatsNonSse()` functions take significant and comparable time (Table 2). When detecting several classes, the run time distribution changes. `CalculateLeafValuesMulti()` calculations (Table 3) take more significant time. If a complex feature extraction algorithm is used, the profile changes significantly. The feature extraction algorithm takes the first place in terms of run time (Table 4).

**Table 2.** Profiling results of CatBoost prediction on the `YearPredictionMSD` dataset on RISC-V CPU. The code was run in a serial mode. Time is given in seconds.

| Function/metric | Call count | Baseline | | Optimized | | Speedup |
|---|---|---|---|---|---|---|
| | | time | % total time | time | % total time | |
| CalcTreesBlockedImpl | 8 | 1.35 | 89.41% | 0.39 | 79.82% | **3.43** |
| CalcIndexesBasic | 79992 | 1.02 | 67.60% | 0.07 | 15.11% | **13.68** |
| CalculateLeafValues | 79992 | 0.21 | 13.70% | 0.20 | 40.56% | **1.03** |
| BinarizeFloatsNonSse | 720 | 0.09 | 5.63% | 0.03 | 6.05% | **2.85** |
| Other (profiler, auxiliary func ...) | | 0.07 | 4.95% | 0.07 | 14.14% | - |
| **Total time** | | **1.51** | | **0.49** | | **3.06** |

**Table 3.** Profiling results of CatBoost prediction on the `Covertype` dataset on RISC-V CPU. The code was run in a serial mode. Time is given in seconds.

| Function/metric | Call count | Baseline | | Optimized | | Speedup |
|---|---|---|---|---|---|---|
| | | time | % total time | time | % total time | |
| CalcTreesBlockedImpl | 8 | 1.45 | 95.70% | 0.81 | 93.17% | **1.79** |
| CalcIndexesBasic | 39520 | 0.70 | 46.40% | 0.06 | 6.33% | **12.76** |
| CalculateLeafValuesMulti | 39520 | 0.67 | 44.44% | 0.69 | 78.64% | **0.98** |
| BinarizeFloatsNonSse | 432 | 0.02 | 1.24% | 0.01 | 1.49% | **1.45** |
| Other (profiler, auxiliary func ...) | | 0.05 | 3.05% | 0.05 | 5.34% | - |
| **Total time** | | **1.52** | | **0.87** | | **1.74** |

In Tables 2-4 we show the computation time of the baseline version of CatBoost (the "*Baseline" column*) and the vectorized version (the "*Optimized" column*). Based on the presented results, it can be seen that the `CalcIndexesBasic()` function has been accelerated by an order of magnitude. The high speedup is due both to the use of vector extensions and to a reduction in the number of operations performed. The speedup of the `BinarizeFloatsNonSse()` function ranged from 1.45 to 5.44 times. Using embedding to extract features (Table 4), resulted in a speedup greater than 3.5 times due to the vector calculation of the L2 norm. However, the

`CalculateLeafValues()` and `CalculateLeafValuesMulti()` functions were not changed due to the lack of efficient vectorization capabilities.

Finally, due to the optimizations performed, it was possible to achieve a speedup of the prediction phase from 1.8 times to 3.7 times. When performing experiments (Tables 2-4), we compared the numerical data with the results obtained on x86 CPUs. The average deviation did not exceed $10^{-11}$. The difference may be due to the order of calculations and is not significant, because the prediction results are the same.

**Table 4.** Profiling results of CatBoost prediction on the `image-embedding` dataset on RISC-V CPU. The code was run in a serial mode. Time is given in seconds.

| Function/metric | Call count | Baseline | | Optimized | | Speedup |
|---|---|---|---|---|---|---|
| | | time | % total time | time | % total time | |
| CalcTreesBlockedImpl | 8 | 1.60 | 8.15% | 1.17 | 18.54% | **1.36** |
| CalcIndexesBasic | 38064 | 0.35 | 1.76% | 0.04 | 0.56% | **9.72** |
| CalculateLeafValues Multi | 38064 | 1.18 | 6.05% | 1.07 | 16.95% | **1,11** |
| BinarizeFeatures | 1 | 17.93 | 91.60% | 5.10 | 80.70% | **3.51** |
| BinarizeFloats NonSse | 312 | 0.03 | 0.13% | 0.00 | 0.07% | **5.44** |
| embeddingProcessing Collection | | 17.91 | 91.48% | 5.10 | 80.63% | **3.51** |
| Other (profiler, auxiliary func ...) | | 0.05 | 0.25% | 0.05 | 0.76% | - |
| **Total time** | | **19.58** | | **6.33** | | **3.10** |

**Table 5.** Final comparison results. The code was run in a multithreaded mode. Time is given in seconds. An accuracy is same in all runs, therefore it is shown only once for each dataset.

| DataSet | Accuracy | Time (x86) | Time (RISC-V) Baseline | Time (RISC-V) Optimized | Speedup |
|---|---|---|---|---|---|
| Santander customer transaction | 0.911 | 0.17 | 16.07 | 7.65 | **2.10** |
| Covertype | 0.960 | 0.42 | 59.41 | 30.60 | **1.94** |
| YearPredictionMSD | 9.168 | 0.06 | 16.30 | 2.79 | **5.84** |
| MQ2008 | 0.850 | 0.02 | 0.55 | 0.50 | **1.10** |
| image-embeddings | 0.802 | 0.18 | 16.66 | 6.00 | **2.78** |

### 5.3 Performance and Accuracy on Full Datasets

The results are summarized in Table 5. Datasets from Table 1 were used for testing. The achieved accuracy was compared between implementations on x86 CPU and RISC-V CPU both for the baseline and the optimized implementations. The metric values coincided, which confirms the correctness of the optimization. Performance was assessed based on full datasets, with each run being performed multiple times, and the average computation times being recorded. The running time on the x86 server is provided for reference purposes only. The code optimization results show significant acceleration for most datasets.

We would also like to highlight two possible limitations of using our results. Firstly, the speedup is achieved only for the use case where prediction is carried out on a batch of samples. When using a single sample, no gain is typically expected. Secondly, we have shown that when solving different types of problems using CatBoost, the hotspots and distribution of computational load among them may vary. We have explored various models, but it is possible that in some other scenarios, further optimization may be required to effectively utilize the resources of RISC-V CPUs.

## 6    Conclusion

The RISC-V ecosystem is evolving at a rapid pace. The current level of infrastructure development allows porting of state-of-the-art software onto existing low-power RISC-V devices, as well as identifying the most promising RISC-V-specific approaches to improve performance. In this paper, we summarized our experience of porting the high-performance CatBoost library, which implements gradient boosting of decision trees, to the Lichee Pi 4A device. We found that large enough code can be recompiled to RISC-V with relative ease and the results are correct. On our path to achieving better performance, we encountered a number of obstacles. First, current compilers can only support vectorization of loops for the RVV 1.0 vector extensions, while almost all RISC-V CPUs use the RVV 0.7.1 extension. Secondly, the available profiling tools have limited capabilities compared to their x86 counterparts. To overcome these problems, we manually profiled the code using simple custom timing tools and identified the main hotspots and critical paths in the call graphs. Next, we vectorized the hotspots using intrinsics and achieved a speedup (up to 5.8 times) in the case when samples from datasets are sent for prediction in batches. We hope that our experience can be used to port other frameworks to RISC-V boards. The code, data, and experimental setups to reproduce the results are available on GitHub [29].